\documentclass[10pt,twocolumn,showpacs,amsmath,amssymb,floatfix,superscriptaddress]{revtex4-2}
\usepackage{graphicx}
\usepackage{float}
\usepackage{dcolumn}
\usepackage{bm}
\usepackage{color}
\usepackage{txfonts}
\usepackage{microtype}
\usepackage{hyperref}
\usepackage{mathrsfs}
\usepackage{easyReview}
\usepackage{amsmath}
\usepackage{extarrows}
\usepackage{mathtools}
\usepackage{xcolor}
\usepackage{slashed}
\usepackage{amsmath}
\usepackage{upgreek}

\setreviewson
\begin{document}

\title{Generation of $\gamma$ photons with extremely large orbital angular momenta }

\author{Ren-Tong Guo}
\affiliation{Ministry of Education Key Laboratory for Nonequilibrium Synthesis and Modulation of Condensed Matter, Shaanxi Province Key Laboratory of Quantum Information and Quantum Optoelectronic Devices, School of Physics, Xi'an Jiaotong University, Xi'an 710049, China}

\author{Mamutjan Ababekri}\email{mamutjan@xjtu.edu.cn}
\affiliation{Ministry of Education Key Laboratory for Nonequilibrium Synthesis and Modulation of Condensed Matter, Shaanxi Province Key Laboratory of Quantum Information and Quantum Optoelectronic Devices, School of Physics, Xi'an Jiaotong University, Xi'an 710049, China}

\author{Qian Zhao}
\affiliation{Ministry of Education Key Laboratory for Nonequilibrium Synthesis and Modulation of Condensed Matter, Shaanxi Province Key Laboratory of Quantum Information and Quantum Optoelectronic Devices, School of Physics, Xi'an Jiaotong University, Xi'an 710049, China}

\author{Yousef I. Salamin}
\affiliation{Department of Physics, American University of Sharjah, Sharjah, POB 26666 Sharjah,  United Arab Emirates}

\author{Liang-Liang Ji}
\affiliation{State Key Laboratory of High Field Laser Physics and CAS Center for Excellence in Ultra-intense Laser Science,Shanghai Institute of Optics and Fine Mechanics (SIOM), Chinese Academy of Sciences (CAS), Shanghai 201800, China}

\author{Zhi-Gang Bu}
\affiliation{State Key Laboratory of High Field Laser Physics and CAS Center for Excellence in Ultra-intense Laser Science,Shanghai Institute of Optics and Fine Mechanics (SIOM), Chinese Academy of Sciences (CAS), Shanghai 201800, China}

\author{Zhong-Feng Xu}
\affiliation{Ministry of Education Key Laboratory for Nonequilibrium Synthesis and Modulation of Condensed Matter, Shaanxi Province Key Laboratory of Quantum Information and Quantum Optoelectronic Devices, School of Physics, Xi'an Jiaotong University, Xi'an 710049, China}

\author{Xiu-Feng Weng}\email{nandswk@163.com}
\affiliation{National Key Laboratory of Intense Pulsed Radiation Simulation and Effect}
\affiliation{Northwest Institute of Nuclear Technology, Xi'an,710024,China}

\author{Jian-Xing Li}\email{jianxing@xjtu.edu.cn}
\affiliation{Ministry of Education Key Laboratory for Nonequilibrium Synthesis and Modulation of Condensed Matter, Shaanxi Province Key Laboratory of Quantum Information and Quantum Optoelectronic Devices, School of Physics, Xi'an Jiaotong University, Xi'an 710049, China}
\affiliation{Department of Nuclear Physics, China Institute of Atomic Energy, P. O. Box 275(7), Beijing 102413, China}
\date{\today}

\begin{abstract}
 Vortex $\gamma$ photons, which carry large intrinsic orbital angular momenta (OAM), have significant applications in nuclear, atomic, hadron, particle and  astro-physics, but their production remains unclear. In this work, we investigate the generation of such photons from nonlinear Compton scattering of circularly polarized monochromatic lasers on vortex electrons. We develop a quantum radiation theory for ultrarelativistic vortex electrons in lasers by using the harmonics expansion and spin eigenfunctions, which allows us to explore the kinematical characteristics, angular momentum transfer mechanisms, and formation conditions of vortex $\gamma$ photons. The multiphoton absorption of electrons enables the vortex $\gamma$ photons, with fixed polarizations and energies, to exist in mixed states comprised of multiple harmonics. Each harmonic represents a vortex eigenmode and has transverse momentum broadening due to transverse momenta of the vortex electrons. The large topological charges associated with vortex electrons offer the possibility for $\gamma$ photons to carry adjustable
 OAM quantum numbers from tens to thousands of units, even at moderate laser intensities.  $\gamma$ photons with large OAM and transverse coherence length can assist in influencing quantum selection rules and extracting phase of the scattering amplitude in scattering processes.

\end{abstract}
\maketitle

Vortex particles are characterized as non-plane-wave states of particles that possess nonzero orbital angular momenta (OAM) relative to their average propagation directions \cite{allen1992orbital,yao2011orbital,knyazev2018beams,bliokh2017theory}. These states represent discrete high dimensional quantum systems, which offer the potential to encode information beyond the one-bit per photon used in quantum teleportation \cite{erhard2018twisted}. When a single photon carrying large intrinsic OAM and reaching up to MeV energy interacts with matter, it induces strong angular momentum effects. These effects manifest themselves as high-multipole excitations of hydrogen-like atoms near a phase singularity, which results in modification of the quantum selection rules \cite{Afanasev2013off,Afanasev2016High,Afanasev2018Experimental}. Additionally, the absorption of higher-order twisted photons can manipulate the rotation of the nuclei \cite{harakeh2001giant,jentschura2011compton}, while the excitation of atoms into high-lying circular Rydberg states presents new venues for studying photo-effects and ionization of atoms \cite{jentschura2011generation,j2015control}. Although optical vortex photons can be generated using techniques such as fork holograms or spiral phase masks \cite{arlt1998production}, assigning large intrinsic OAM  quantum numbers, ranging from tens to thousands, to a $\gamma$ photon remains a challenge.

OAM of a single $\gamma$ photon can be obtained through the angular momentum transfer mechanisms during particle scattering or radiation processes \cite{ivanov2011colliding,ivanov2011scattering,ivanov2022promises,sun2022}, in which the spin angular momenta (SAM) and OAM of initial particles are transferred to OAM of final particles. Photons in the radiation filed of the $n$-th harmonic scattered from high-energy electrons in laser or undulator fields carry intrinsic OAM in the amount $(n-1)\hbar$, where $n\hbar$ is SAM of the absorbed photons \cite{sasaki2008proposal,Bahrdt:2013eoa,katoh2017angular,taira2017gamma,bogdanov2019semiclassical,bogdanov2020planar,kazinski2021radiation,karlovets2022generation,Karlovets:2022mhb,ababekri2023vortex}. To fully describe the evolution of electrons and reveal results that are missing from classical pictures, a treatment within the context of quantum electrodynamics is necessary \cite{karlovets2022generation,Karlovets:2022mhb,ababekri2023vortex}. In cases where laser intensities are weak, the production of higher harmonics is less likely. As the laser intensity increases, multiple emissions of radiation by electrons can destroy the initial coherent state, and lead to a conical distribution of radiated photons, thus complicating formation conditions of the vortex $\gamma$ photons \cite{ivanov2011colliding,ivanov2011scattering,karlovets2022generation,Karlovets:2022mhb,ababekri2023vortex}. Both of these aspects reduce the transfer efficiency of SAM to large intrinsic OAM. On the other hand, vortex lasers or photons enable the transfer of OAM by resonant elastic scattering of relativistic ions \cite{karlovets2021wave} or Compton scattering of electrons \cite{sasaki2008proposal,Bahrdt:2013eoa,katoh2017angular,taira2017gamma,bogdanov2019semiclassical,bogdanov2020planar,kazinski2021radiation, jentschura2011generation,jentschura2011compton,petrillo2016compton,maruyama2019compton}.   On account of the threshold limitation on phase plate materials, the topological charges of intense vortex lasers do not generally exceed ten \cite{yao2011orbital}. Although the production of large OAM $\gamma$ photons via nonlinear Compton scattering (NCS) is widely reported  \cite{gong2018brilliant,liu2020vortex,wang2020generation,hu2021attosecond,zhang2021efficient,bake2022bright,  chen2018gamma,chen2019generation}, OAM of whole beams are collective features rather than a single-photon property. Nevertheless, relying solely on vortex lasers as sources of large intrinsic OAM is not sufficient, and the  yield and brightness of NCS can be higher than in the corresponding linear process \cite{bula1996observation}. In principle, vortex electrons can serve as a source of OAM and through NCS make it possible to generate $\gamma$ photons with large intrinsic OAM. 

At present, vortex electron beams are generated experimentally by using  nanofabricated diffraction holograms \cite{verbeeck2010production,mcmorran2011electron}. Their OAM can reach up to about 1000$\hbar$ \cite{Mafakheri2017realization}, with kinetic energies in the range of 200-300 keV. The beam energy can be boosted to MeV or GeV levels by injection into a homogeneous and stable linear accelerator \cite{ivanov2011scattering,  ivanov2022promises}, or a cyclotron \cite{Karlovets2021tlg,silenko2017manipulating, silenko2019siberian,lloyd2017electron,bliokh2017theory,bliokh2007semiclassical}, to avoid depolarization of the resonant orbitals. Thanks to the transverse momenta and large topological charges of vortex electrons, the associated radiation processes get spatial distributions different from those of the plane-wave states \cite{kaminer2016quantum,ivanov2016quantum,ivanov2013detecting,pupasov2022passage,karlovets2021nonlinear,wang2022triple,groshev2020bremsstrahlung,seipt2014structured}.  However, studies on laser interaction with vortex electrons focus more on the electron dynamics, than on the radiation aspects \cite{karlovets2012electron,hayrapetyan2014interaction,bandyopadhyay2015relativistic,aleksandrov2022scattering}. Interaction with rotationally invariant systems, such as circularly polarized (CP)  lasers, can result in complete SAM transfer. It is crucial to elucidate angular momentum transfer mechanisms when vortex electrons absorb multiple laser photons from CP lasers. Effects of the transverse momenta of vortex electrons on the formation conditions and properties of vortex $\gamma$ photons are not entirely clear. A more comprehensive quantum treatment of this problem is necessary for NCS on ultrarelativistic vortex electrons \cite{ababekri2023vortex}. 

\begin{figure}[!t]
	\setlength{\abovecaptionskip}{-0.0cm}
	\includegraphics[width=0.9\linewidth]{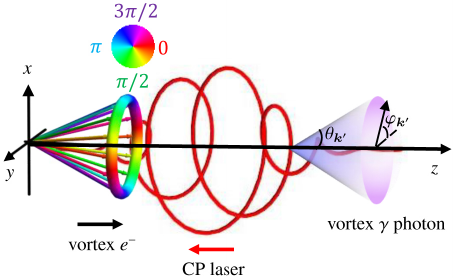}
\vskip-0.0cm	
\caption{A schematic illustrating generation of $\gamma$ photons with large intrinsic OAM during NCS of the ultrarelativistic vortex electron in a CP monochromatic laser. The vortex states of the initial electron and emitted photon are represented in momentum space by cones with cone angles $\theta_{\bm{q}}=\arctan\frac{|\bm{q}_\perp|}{q_z}, \theta_{\bm{k}^{\prime}}=\arctan\frac{|\bm{k}^{\prime}_\perp|}{k_z}$ and azimuthal angles $\varphi_{\bm{q}},\varphi_{\bm{k}^{\prime}}$, relative to the momentum vectors $\bm{q}$ and $\bm{k}^\prime$, respectively. }
	\label{fig_illustration}
\end{figure}

\begin{figure}[!t]
	\setlength{\abovecaptionskip}{-0.0cm}
	\includegraphics[width=.95\linewidth]{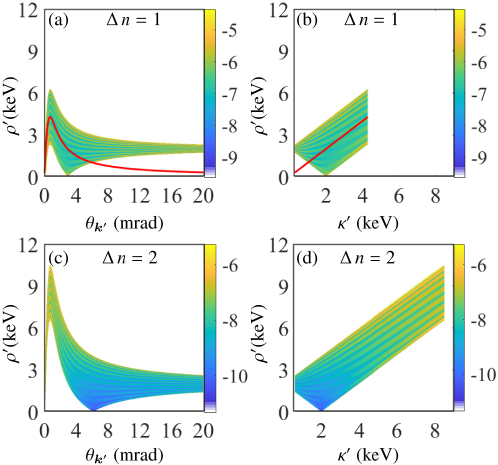}

	\caption{Top and bottom rows pertain to vortex electrons with $m_{\rho}=10$ absorbing one and two laser photons, respectively. Probabilities of radiating vortex $\gamma$ photons in (a) and (c): $\log_{10}[d^{2}W/(d\rho^{\prime}d\theta_{\bm{k}^{\prime}})]$ and in (b) and (d): $\log_{10}[d^{2}W/(d\rho^{\prime}d\kappa^{\prime})]$. The red lines represent the case when the initial electron state is plane-wave.}
	\label{fig_2}
\end{figure}

\begin{figure}[!t]
	\setlength{\abovecaptionskip}{-0.0cm}
	\includegraphics[width=1\linewidth]{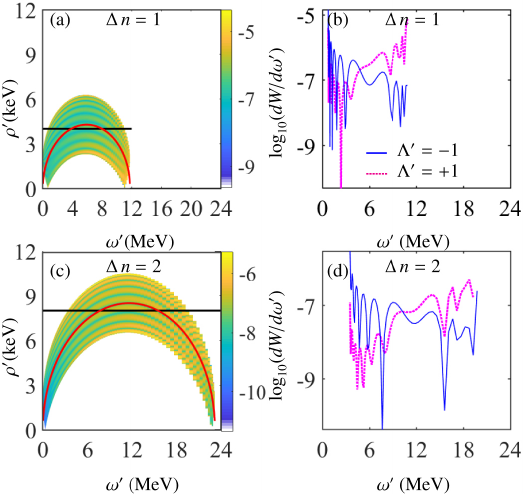}

	\caption{Top and bottom rows pertain to vortex electrons with $m_{\rho}=10$ absorbing one and two laser photons, respectively. Probabilities of radiating vortex $\gamma$ photons in (a) and (c): $\log_{10}[d^{2}W/(d\rho^{\prime}d\omega^{\prime})]$. Red lines represent the case when the initial electron state is a plane-wave. Black lines fix values of the final electron transverse momenta randomly ($\rho^\prime=4$ and $8$ keV). (b) and (d): Energy spectra of $\gamma$ photon with  polarizations $\Lambda^\prime=\pm1$ when $\rho^\prime$ is fixed, respectively. Blue and magenta dotted lines represent $\Lambda^\prime=-1$ and $\Lambda^\prime=1$, respectively.}
	\label{fig_3}
\end{figure}

\begin{figure}[!t]
	\setlength{\abovecaptionskip}{-0.0cm}
	\includegraphics[width=1.0\linewidth]{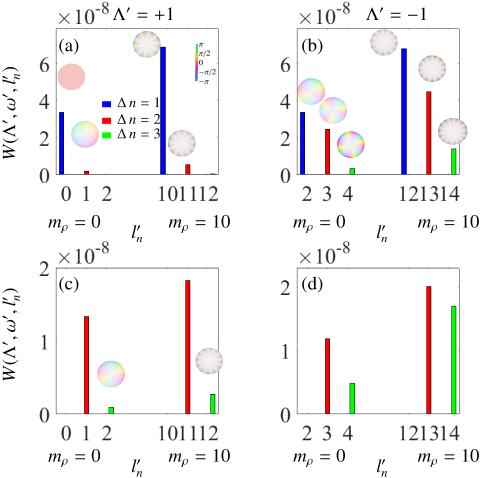}
	
\vskip0.2cm	
	\caption{Top and bottom rows are the OAM and phase distributions of vortex $\gamma$ photons with fixed energies $\omega^{\prime}=6$ MeV and 12 MeV, respectively. (a) and (c): $\Lambda^{\prime}= +1$;  (b) and (d): $\Lambda^{\prime}=- 1$. Note that the OAM values start from $l_1^\prime=0$ for $\Lambda^\prime=+1$ and $l_1^\prime=2$ for $\Lambda^\prime=-1$ because of $l^\prime_n=\Delta\,n-\Lambda^\prime$. The blue, red and green columns represent absorption of one, two and three laser photons, respectively. Left and right halves of each graph represent $m_\rho=0$ and $m_\rho=10$, respectively.} 
	\label{fig_4}
\end{figure}

In this Letter, we investigate the generation of $\gamma$ photons with large intrinsic OAM during the NCS on ultrarelativistic vortex electrons. We develop a fully quantum radiation theory of vortex electrons in intense CP monochromatic lasers, to elucidate the kinematical features, angular momentum transfer mechanisms and formation conditions of vortex $\gamma$ photons. The collision geometry is presented in Fig.~\ref{fig_illustration}, where the initial electron is in a vortex Bessel state propagating along $z-$axis to collide with a counter-propagating CP laser. We find that the conservation of angular momentum is obtained when all particles are coaxial, while the introduction of transverse wavepackets to initial vortex electrons makes it possible to generate vortex $\gamma$ photons slightly off-axis and obtain an approximate angular momentum transfer relation. Consequently, $\gamma$ photons with controllable OAM quantum numbers from tens to thousands can be generated through OAM  transfer from vortex electrons at moderate laser intensities. Each harmonic exhibits transverse momentum broadening, owing to the transverse momenta of vortex electrons (see Figs.~\ref{fig_2} and \ref{fig_3}). The electron absorbs multiple laser photons and emits vortex $\gamma$ photons with fixed energies and polarizations in incoherent mixed states consisting of different harmonics (see Figs.~\ref{fig_3} and \ref{fig_4}). Natural units are used ($\hbar = c = 1$) and $\alpha=\frac{e^2}{4 \pi}$ is the fine structure constant, where $e$ is the electron charge. The momentum four-vector will be denoted by $q=(\varepsilon, \bm{q})$, where  $\varepsilon$ is the energy and $\bm{q}=(\bm{q}_\perp, q_z)=(|\bm{q}_\perp|, \varphi_{\bm{q}}, q_z)$ is the 3-momentum vector, with $\bm{q}_\perp$ and $q_z$ its transverse and longitudinal components, respectively.

We first explore the angular momentum transfer mechanism of vortex $\gamma$ photons on the $z-$axis. In the calculation, all particles carry SAM and OAM, and the vortex modes will be represented by Bessel functions for simplicity, thus the $S$-matrix element of a vortex electron radiating vortex $\gamma$ photons in the interaction with a CP monochromatic laser is \cite{supplemental}:
\begin{eqnarray}
	\begin{aligned}
		&\ S_{fi}^{vortex}=\langle \rho^{\prime},m_{\rho^\prime},q^{\prime}_{z}, \lambda^{\prime} ; \kappa^{\prime},m_{\kappa^{\prime}}, k_{z}^{\prime}, \Lambda^{\prime}\vert \hat{S} \vert\rho,m_{\rho}, q_{z},\lambda,\rangle \\ 
		&=\mathcal{N}_{NCS}\delta\left(\sum_i \varepsilon_{i}\right) \delta\left(\sum_i q_{z, i}\right)\sum_{\sigma,\sigma_{\lambda},\sigma_{\lambda^\prime}}^{n_{i},n_{f},\sigma_{\Lambda^\prime}} \delta_{\tilde{m}_{\rho}-m_{\rho},\tilde{m}_{\kappa^{\prime}}-m_{\kappa^{\prime}}+\tilde{m}_{\rho^\prime}-m_{\rho^\prime}} \\ 
		&\times \mathscr{M}(\sigma, \sigma_{\lambda},\sigma_{\lambda^\prime},\sigma_{\Lambda^\prime},n_{i},n_{f}),
	\end{aligned}
\label{S_fi_3vortex}
\end{eqnarray}
where $\mathcal{N}_{NCS}=\frac{\alpha}{16 \pi \varepsilon [(q+\Delta nk) \cdot k^\prime]}$, $\delta(\sum_i \varepsilon_{i})$ and $\delta(\sum_i q_{z, i})$ are shorthands for  conservation of energy and momentum in the $z$ direction, respectively. $\tilde{m}_{\rho}=n_i+\sigma_{\lambda}-\lambda, \tilde{m}_{\rho^{\prime}}=n_f-\sigma+\sigma_{\lambda^\prime}-\lambda^\prime, \tilde{m}_{\kappa^{\prime}}=\sigma_{\Lambda^\prime}$, and $n_i\ (n_f)$ and $\lambda\ (\lambda^\prime)$ are the harmonic indicator and spin of the initial (final) electrons, respectively. In order to completely extract the azimuthal information, it is necessary to expand the Volkov states, bi-spinors and polarization vectors, and introduce the dummy indices $n_i, n_f, \sigma, \sigma_\lambda, \sigma_{\lambda^\prime}, \sigma_{\Lambda^\prime}$ \cite{varshalovich1988quantum}. Other parameters are defined in \cite{supplemental}. We use
$\vert\rho,m_{\rho}, q_{z},\lambda\rangle=\int \frac{d^{2}\bm{q}_{\perp}}{(2\pi)^{2}} (-i)^{m_{\rho}}e^{im_{\rho}\varphi_{\bm{q}}}\sqrt{\frac{2\pi}{\rho}}\delta(|\bm{q}_{\perp}|-\rho)\Psi_{q,\lambda}$ to represent the incoming electron,  where $m_{\rho}$ and $|\bm{q}_{\perp}|=\rho$ are the topological charge and the modulus of transverse momentum, respectively. The outgoing electron and emitted photon are represented by these similar forms $\langle\rho^{\prime},m_{\rho^\prime},q^{\prime}_{z}, \lambda^{\prime}\vert$ and $\langle\kappa^{\prime},m_{\kappa^{\prime}}, k_{z}^{\prime}, \Lambda^{\prime}\vert$ with the topological charges $m_{\rho^\prime}, m_{\kappa^\prime}$ and the moduli of transverse momenta $|\bm{q}^{\prime}_{\perp}|=\rho^{\prime}, |\bm{k}^{\prime}_{\perp}|=\kappa^{\prime}$, respectively. Note that the explicit kinematical relation in the transverse direction $\delta^{2}(\bm{q}_{\perp}-\bm{q}_{\perp}^{\prime}-\bm{k}_{\perp}^{\prime})$ is included in the amplitude $\mathscr{M}$ \cite{supplemental}. The allowed kinematical region of $|\bm{q}_{\perp}^{\prime}|$ and $|\bm{k}_{\perp}^{\prime}|$ for a fixed $\rho$ is defined by the ‘‘triangle rules’’. Integrals in $\mathscr{M}$ come only from two points in the entire ($\bm{q}_{\perp}^{\prime}, \bm{k}_{\perp}^{\prime}$) space \cite{ivanov2011colliding,ivanov2011scattering,ivanov2022promises,Ivanov2020kinematic}. We can associate these two configurations with  two different paths from the initial state to the final state in momentum space, which explains the appearance of interference fringes in Figs.~\ref{fig_2} and \ref{fig_3}.  If the initial electron state is a plane-wave, the transverse momenta of final particles correspond one by one and the above phenomena disappear.

The probability of a vortex electron radiating vortex $\gamma$ photons in the interaction with a CP monochromatic laser is \cite{supplemental}:
\begin{eqnarray}
	\begin{aligned}
		\frac{d^2W_{vortex}}{d\rho^{\prime} d\theta_{\bm{k}^{\prime}} \Delta m_{\kappa^{\prime}}}=\mathcal{N}_{NCS}\rho \rho^\prime\kappa^\prime \pi\sum_{\Delta n}\delta_{m_{\kappa^\prime},\, m_{\rho}+\Delta n+\lambda-m_{\rho^\prime}-\lambda^{\prime}}|\mathscr{M}|^{2},
	\end{aligned}
\label{rate_vortex}
\end{eqnarray}
after eliminating the dummy indices $\{\sigma, \sigma_{\lambda}, \sigma_{\lambda^\prime},\sigma_{\Lambda^\prime}\}$ as in \cite{supplemental}. This leads to the conclusion that the angular momentum is conserved via $\delta_{m_{\kappa^\prime},\, m_{\rho}+\Delta n+\lambda-m_{\rho^\prime}-\lambda^{\prime}}$ when all particles are coaxial, where $\Delta\,n$ represents the harmonic number. We underline that if the electron radiates multiple times, the emitted photons will be conical in distribution and no longer meet the coaxial conditions of all particles, which limits the laser intensity to moderate values ($\xi\sim1$). Since the spin flip ($\lambda\neq\lambda^\prime$) probability is negligible under moderate laser intensities \cite{ababekri2023vortex}, a simpler form emerges, namely  $\delta_{m_{\kappa^\prime},\Delta\,n+m_{\rho}-m_{\rho^\prime}}$ and each harmonic represents a vortex eigenmode for the emitted $\gamma$ photon. OAM of the $\gamma$ photon can be precisely regulated by OAM of vortex electrons and SAM of absorbed laser photons. Since the probability of radiation decreases with the increasing harmonic number $\Delta\,n$, OAM of the vortex electrons $m_{\rho}$ must be the main source of the large OAM of the $\gamma$ photons.

Radiation from the vortex electron can be investigated numerically, taking the electron energy 
 $\varepsilon=1$ GeV and the cone angle $\theta_{\bm{q}}=2\times 10^{-3}$mrad (corresponding to $\rho=2\, \text{keV}$). Such electrons can be produced in circular storage rings and linear accelerators to boost the low energy vortex electrons \cite{lloyd2017electron,bliokh2017theory,bliokh2007semiclassical,Karlovets2021tlg}. In our investigations, we assume that all angular momentum is transferred to the final $\gamma$ photon ($m_{\rho^\prime}=0$), the other OAM cases for the final particles are shown in \cite{supplemental}, and the topological charge of the initial electron is 10 (they can reach about 1000 in current experiments \cite{Mafakheri2017realization}). Furthermore, we assume laser intensity $\xi=1$ and photon energy $\omega=1.55$ eV (corresponding to an optical laser of intensity $I\approx 2\times 10^{18}$ W/cm$^{2}$ and wavelength  $\lambda_0=0.8\, \upmu$m, available at few-hundred-terawatt facilities.) Modern laser facilities can achieve peak intensities in excess of $10^{22}$ W/cm$^{2}$ \cite{danson2019petawatt,cartlidge2018light,kawanaka2016conceptual,Yoon:21}.

Figure \ref{fig_2} shows radiation probability distributions of the vortex electron in the $(\theta_{\bm{k}^{\prime}}, \rho^\prime)$ and $(\kappa^\prime, \rho^\prime)$ planes, where $\theta_{\bm{k}^{\prime}}, \kappa^\prime$ and $\rho^\prime$ are the cone angle of the emitted photon, and the moduli of transverse momenta of the emitted photon and outgoing electron, respectively. Since vortex electrons possess transverse momenta, kinematic areas are broad compared to the case of employing plane-wave electrons [see the red lines in Figs.~\ref{fig_2} (a) and (b)]. We find that, when $m_{\rho}=10$, the interference fringes are obvious. The fringes stem from integrals in $\mathscr{M}$ \cite{supplemental}, each of integrals can be represented by $\int drrJ_{m^\prime}(\kappa^\prime r)J_{m-m^\prime}(\rho^\prime r)J_{m}(\rho r)=\cos(m^\prime \angle _{\kappa^\prime,\,\rho}-(m-m^\prime)\angle_{\rho,\,\rho^\prime})/(2\pi A_{\kappa^\prime,\,\rho^\prime,\,\rho})$, where $A_{\kappa^\prime,\,\rho^\prime,\,\rho}$ is the area of a triangle made up of transverse momenta of all particles \cite{jackson1972integrals}. The angles of the triangle ($\angle _{\kappa^\prime,\,\rho}, \angle_{\rho,\,\rho^\prime}$) vary with the transverse momenta, causing oscillations in the probability distributions. The integrals contain contributions from two kinematic configurations in the transverse plane and include conservations of both transverse and angular momenta. Note that the intersection values of the transverse or longitudinal axes in the $(\kappa^\prime, \rho^\prime)$ planes and the number of fringes are manifestations of the transverse momenta and topological charges, respectively, of initial vortex electrons \cite{ivanov2011colliding,ivanov2011scattering,ivanov2022promises,Ivanov2020kinematic}. 

Energy spectra $\omega^{\prime}$ of vortex $\gamma$ photons are displayed in Fig.~\ref{fig_3}. The broadening and interference fringes caused by transverse momenta of the vortex electrons are still clearly visible in the ($\omega^\prime$, $\rho^\prime$) planes. Each group of coordinates ($\rho^{\prime},\Lambda^{\prime}, \omega^{\prime}$) represents an independent radiation
event and defines a coherent vortex state with definite OAM, hence the choice of fixed $\rho^\prime=4$ and $8$ keV for the convenience of analysis [see the black lines in Figs.~\ref{fig_3} (a) and (c)]. Energy spectra for fixed $\rho^\prime$ and polarization $\Lambda^{\prime}$ oscillate up and down as they pass through the light and dark fringes. Clearly, different harmanics with $\Lambda^{\prime}=+1$ dominate in the higher energy region because the polarization of the laser photon is $+1$ [see Figs.~\ref{fig_3} (b) and (d)]. Different harmonics commonly contribute to the low-energy region, so the vortex $\gamma$ photons are in mixed states. Energy truncation of the lower harmonics and discontinuities in the higher harmonics spectra in monochromatic lasers, allow vortex $\gamma$ photons with only one harmonic contribution to be  implemented by filtering appropriate energy ranges. Therefore, when  $\rho^\prime$ is fixed, vortex $\gamma$ photons of different $\omega^{\prime}$ will be in pure or mixed states. 

OAM properties of vortex $\gamma$ photons with fixed energies $\omega^\prime=6$ Mev or 12 MeV and polarizations $\Lambda^{\prime}=\pm1$ radiated from plane-wave electrons ($m_{\rho}=0$) and vortex electrons ($m_{\rho}=10$) are presented in Fig.~\ref{fig_4}. Large OAM $l^\prime_n$ of $\gamma$ photons can be achieved from either the harmonic number $\Delta\,n$ or the topological charge of incoming electron $m_{\rho}$ due to $l^\prime_n=m_{\rho}+\Delta\,n -\Lambda^\prime$. The probability of $\Delta\,n=2$ is already one order of magnitude smaller than that of $\Delta\,n=1$, so it is more likely to depend on $m_{\rho}$. Once their energies and polarizations are determined, vortex $\gamma$ photons are in incoherent mixed states composed of multiple harmonics. This is a result of the electrons, both plane-wave and vortex states, absorbing multiple laser photons. While the harmonics radiated by vortex electrons contain transverse momentum broadening in $\kappa^\prime$. In addition, we find that due to the harmonic energy truncation in monochromatic lasers, when a vortex $\gamma$ photon is at a higher energy $\omega^\prime=12$ MeV [see Figs.~\ref{fig_4} (c) and (d)], there is no contribution from the main harmonic. In pulsed CP lasers, however, contribution from the main harmonic always exists, because of ponderomotive broadening of harmonics. Due to the difference in energy between the laser photons in the pulse and the central laser photon, $\kappa^\prime$ broadening of vortex $\gamma$ photons occurs even in the case of an initial plane-wave electron, and yet OAM of $\gamma$ photons is derived only from $\Delta\,n$ \cite{ababekri2023vortex}. Pulse, polarization and focusing effects of real lasers \cite{tang2020highly,mackenroth2011nonlinear} make vortex $\gamma$ photons different in phase structure and $\kappa^{\prime}$ broadening.

We investigate effects of the laser and electron parameters on the vortex $\gamma$ photons in \cite{supplemental}. Other parameters do not affect the above analysis, but change the energy ranges, transverse momentum distributions, harmonics probabilities and interference fringes. When all particles are coaxial, the conservation of angular momentum $\delta_{m_{\kappa^\prime}+m_{\rho^\prime},\,\Delta\,n+m_{\rho}}$ can be obtained.  As for the $m_{\kappa^\prime}$ and $m_{\rho^\prime}$ values, they satisfy the angular momentum conservation, and the radiation dynamics would not be affected by different OAM cases for the final particles \cite{ababekri2023vortex}. The specific vortex states of the final particles can be probed by participating in scattering processes,  such as the collective excitations of different multipole transitions in nuclei \cite{lu2023manipulation} and the Compton sacttering \cite{Sherwin2020effect,Sherwin2017theo,Sherwin2017compton}. The former determines the cone angles and topological charges of $\gamma$ photons by measuring the nuclear photon-absorption cross section, which alters the selection rules and extracts the giant resonances with specific multipolarity \cite{Afanasev2013off,Afanasev2016High,Afanasev2018Experimental}; 
the latter determines phase structures of the vortex $\gamma$ photons by observing the angular distributions of the final particles.  
 
 Let's consider a more common scenario of generating vortex $\gamma$ photons which are slightly off-axis. For a smooth transition of OAM transfer relation from on the $z-$axis to slightly off-axis of the final particles, wave packets with transverse coherence sizes are added to the Bessel modes \cite{ivanov2011colliding,ivanov2011scattering,ababekri2023vortex,Ivanov2011addendum,Ivanov2020kinematic}. This can be accomplished by incorporating $f(\rho)$  into the initial vortex electron and $g(\kappa^{\prime})$ into the final vortex photon. 
The $S$-matrix element becomes: $S_{fi}=\int d\rho d\kappa^{\prime} f(\rho) g(\kappa^{\prime})\frac{d^2\bm{q}_\perp}{(2\pi)^2} \frac{d^2\bm{k}^\prime_\perp}{(2\pi)^2}a_{\rho\,m_{\rho}}(\bm{q}_\perp)a^\ast_{\kappa^\prime\,m_{\kappa^\prime}}(\bm{k}_\perp^\prime)S^{pw}_{fi}\,,$
where $S^{pw}_{fi}$ is the $S-$matrix element for the NCS with plane-wave particles. Thus the master integral becomes \cite{ababekri2023vortex,Ivanov2011addendum}: $\hat{\mathcal{I}}_{ m_{\rho},\, m_{\kappa^\prime}}^{\tilde{m}_\rho\, \tilde{m}_{\kappa^\prime}}(\rho,\kappa^\prime,\bm{Q})=\int d\rho f(\rho) \int d\kappa^\prime g(\kappa^\prime) \mathcal{I}_{ m_{\rho},\,  m_{\kappa^\prime}}^{\tilde{m}_{\rho},\, \tilde{m}_{\kappa^\prime}}(\rho,\kappa^\prime,\bm{Q}),$ where $f(\rho)=N_\rho \exp[- \frac{(\rho-\rho_0)^2}{2 \sigma_\rho}]$ and $g(\kappa^{\prime})=N_{\kappa^{\prime}} \exp[- \frac{(\kappa^{\prime}-\kappa^{\prime}_0)^2}{2 \sigma_{\kappa^{\prime}}}]$ are properly normalized weight functions with peak values at $\rho_0$, $\kappa^{\prime}_0$, the widths $\sigma_\rho$, $\sigma_{\kappa^{\prime}}$ and the normalization constants $N_\rho$, $N_{\kappa^{\prime}}$, respectively. $\hat{\mathcal{I}}_{ m_{\rho},\,  m_{\kappa^\prime}}^{\tilde{m}_\rho,\, \tilde{m}_{\kappa^\prime}}(\rho,\kappa^\prime,\bm{Q})$ includes $\delta^{(3)}(\bm{q} + \Delta\,n \bm{k}-\bm{q}^{\prime}-\bm{k}^{\prime})$ and $\bm{Q}=\bm{q}^\prime - \Delta\,n \bm{k}$.  
It is still possible to obtain the approximate angular momentum transfer relation $\vert m_{\kappa^{\prime}} -(\Delta\,n+m_{\rho}) \vert \lesssim \frac{\rho_0}{\sigma_\rho}$ \cite{ababekri2023vortex,supplemental}, with the approximate equality holding for $\sigma_\rho \gg \rho_0$ in the Gaussian weight function. This proves that, if the initial and final states are wave packets, $m_{\kappa^{\prime}} \approx \Delta\,n+m_{\rho}$ and $\kappa^{\prime} \approx \rho$ during small-angle scattering ($\beta\sim m_e/\varepsilon$, where $m_e$ is the electron mass). Introduction of wave packets is responsible for the very small broadening in the radiation or OAM spectra and smearing of the interference fringes \cite{Ivanov2020kinematic}. Such vortex states are called spatiotemporal vortices, they are unavoidably non-monochromatic and exhibit space and time evolution \cite{bliokh2012spatiotemporal,ivanov2022promises}. Therefore, a scenario in which the initial electron vortex state is prepared as a Gaussian wave packet of suitable transverse coherence size to participate in NCS, can be easier to implement experimentally. Of course, care must be taken to ensure that the laser intensity does not cause multiple emissions which can destroy the initial electron coherent state.  The vortex $\gamma$ photons with wave packets reveal itself in corrections to the conventional cross sections and help to extract the phase of scattering amplitude, which can be additionally enhanced if there are particles with high OAM $l_n^{\prime}\gtrsim10^3$, analogously to collisions at large impact parameters  \cite{maruyama2019compton,Karlovets2020effects,Karlovets2017poss,Karlovets2015gauss,Karlovets2016dva,Karlovets_2017}.

In conclusion, we investigated the production of $\gamma$ photons with large intrinsic OAM via NCS and developed the quantum radiation theory of ultrarelativistic vortex electrons in CP monochromatic lasers. $\gamma$ photons carrying controllable OAM quantum numbers from tens to thousands of units are generated when employing vortex electrons, even at moderate laser intensities. As a consequence of multiphoton absorption by the electrons, vortex $\gamma$ photons of fixed energies and polarizations emerge in incoherent mixed states consisting of different eigenmodes. Each eigenmode has transverse momentum broadening because of the transverse momenta of vortex electrons. This will hopefully lead to a better understanding of the radiation from vortex electrons in laser fields and properties of the emitted vortex $\gamma$ photons, that may open novel opportunities in hadron, atomic, nuclear, particle and high-energy physics, currently unavailable for traditional scattering experiments.

{\it Acknowledgment:}  We thank  I. P. Ivanov for productive discussions. The work is supported by the National Natural Science Foundation of China (Grants No. 12022506, No. U2267204, No. 12105217, No. 12147176, No. 12235013), the Foundation of Science and Technology on Plasma Physics Laboratory (No. JCKYS2021212008), and the Shaanxi Fundamental Science Research Project for Mathematics and Physics (Grant No. 22JSY014).

\bibliography{vblib}

\end{document}